\documentclass[%
 reprint,
 longbibliography,
 amsmath,amssymb,
aps,
prb,
]{revtex4-2}

\usepackage{graphicx}% Include figure files
\usepackage{dcolumn}% Align table columns on decimal point
\usepackage{bm}% bold math
\usepackage{xcolor}% bold math
\begin{document}

\preprint{APS/123-QED}

\title{High pressure and temperature thermoelasticity of hcp osmium from ab initio quasi-harmonic theory}% Force line breaks with \\

\author{Xuejun Gong}
 \affiliation{School of Physical Science and Technology, Xinjiang University, Urumqi, Xinjiang 830046, China}
  \affiliation{International School for Advanced Studies (SISSA), Via Bonomea 265, 34136, Trieste, Italy}
 
\author{Andrea Dal Corso}
\affiliation{International School for Advanced Studies (SISSA), Via Bonomea 265, 34136, Trieste, Italy}
 \affiliation{IOM - CNR, Via Bonomea 265, 34136, Trieste, Italy}
 
\date{\today}% It is always \today, today,

\begin{abstract}

We present a systematic ab initio study of the thermoelastic properties of hcp osmium as functions of temperature and pressure within the quasi-harmonic approximation (QHA). The precision of the Zero Static Internal Stress Approximation (ZSISA) and of the volume-constrained ZSISA (V-ZSISA) are rigorously assessed. For osmium, we find negligible deviations between ZSISA and a full free energy minimization (FFEM) approach. 
Also, the V-ZSISA approximation influences the results very little, as we found already in beryllium, despite the markedly different behavior of the $c/a$ ratio with temperature in the two metals.
Our QHA-derived ECs show excellent agreement with available experimental data in the temperature range of $5–301$ K, outperforming the results obtained from the quasi-static approximation (QSA). Additionally, we report the pressure-dependent QHA ECs at $5$ K, $301$ K, and $1000$ K, spanning pressures from $0$ to $150$ kbar.
\end{abstract}

\maketitle

%\tableofcontents

\section{Introduction}

Osmium, an hexagonal close-packed (hcp) $5d$ transition metal, is notable for its exceptional physical properties, including the highest density among all elements, a remarkably high melting point ($3306$ K), and a bulk modulus comparable to that of diamond. Osmium is also often alloyed with platinum and iridium to enhance their mechanical performance. Its thermodynamic properties have been extensively investigated both experimentally~\cite{arblaster_crystallographic_2013} and theoretically~\cite{liu_structural_2011,palumbo_lattice_2017,burakovsky_ab_2015}. Numerous theoretical studies have focused on its equation of state, structural parameters, and electronic properties under pressure,~\cite{tal_pressure-induced_2016,dubrovinsky_most_2015,koudela_lifshitz_2006} but its elastic properties remain only partially characterized.

The temperature dependence of osmium's elastic constants (ECs) has been experimentally measured at ambient pressure over the range of $5–301$ K~\cite{panteaElasticConstantsOsmium2009c}, and recent work has inferred the pressure dependence of $C_{44}$ (from $0$ to $2022$ kbar) at room temperature from Raman spectroscopy data~\cite{jingyi_liu_high-pressure_2022}. Theoretical calculations of osmium's ECs at ambient pressure and $0$ K have been conducted by several groups using ab initio methods within both the local density approximation (LDA)~\cite{lda} and the Perdew-Burke-Ernzerhof (PBE) generalized gradient approximation.~\cite{pbe,fast_elastic_1995,fan_potential_2006,minisini_elastic_2008,yu_calculations_2010} However, some discrepancies still remain among the reported values. Additionally, the pressure dependence of osmium's ECs has been computed at 0 K~\cite{deng_elastic_2009}, but no theoretical data are available for the temperature-dependent elastic constants (TDECs) within either the quasi-static (QSA) or quasi-harmonic (QHA) approximations. Furthermore, the high-pressure and high-temperature regime remains entirely unexplored, with no measurements or calculations reported to date. \par

In a recent study, we introduced a computational workflow for calculating the TDECs of hcp metals and applied it to beryllium.~\cite{gong_high-temperature_2024} This methodology enables the calculation of both QSA and QHA ECs within the zero static internal stress approximation (ZSISA), where ionic positions are relaxed by minimizing the total energy at each strain. Additionally, the method allows for the quantification of the ZSISA approximation's accuracy by comparing it to results obtained from full free energy minimization (FFEM). Furthermore, the effect of the volume-constrained ZSISA (V-ZSISA) can be assessed by comparing ECs computed along the $0$ kbar isobar with those calculated along the stress-pressure curve (defined as the path in crystal parameters space where the stress is a uniform pressure) at $0$ K. Although the QHA is computationally more demanding than the QSA, it has been demonstrated to provide superior accuracy for several face-centered cubic (fcc) and body-centered cubic (bcc) metals, making it essential for quantitative predictions.~\cite{malica_quasi-harmonic_2020,malica_quasi-harmonic_2021,gong_ab_2024,gong_high-temperature_2024,gong_ta_2025} \par

In this work, we apply this workflow to osmium, presenting a comprehensive analysis of its elastic properties. We compare the ZSISA QHA results with those obtained from FFEM at a selected geometry, and quantify the effects of the V-ZSISA approximation within QSA. QHA TDECs are calculated only within V-ZSISA. We further validate our approach by comparing the QSA and QHA TDECs with experimental data in the temperature range of $5–301$ K and provide predictions for elevated temperatures up to $1600$ K. Additionally, we report the pressure dependence of the ECs from $0$ to $150$ kbar at temperatures of $5$ K, $301$ K, and $1000$ K, offering the first theoretical estimates of osmium's elastic properties in this high-pressure, high-temperature regime. We anticipate that these predictions will serve as a valuable reference for future experimental investigations.\par

\begin{table*}
  \begin{center}
%\begin{minipage}{174pt}
\caption{
Equilibrium lattice constant ($a$), bulk modulus ($B_T$), and first and second pressure derivative of the bulk modulus ($B_T'$, $B_T^{''}$)  for osmium obtained from a fourth-order Birch-Murnaghan fit of the energy. Comparison of current calculations with previous studies and experimental data. At $295$ K we report the adiabatic QHA bulk modulus $B_S$, while at $0$ K $B_S=B_T$. The $1500$ K QHA crystal parameters are also reported (see text).
}\label{table:1}%
\begin{tabular}{@{}llllllll@{}}
\toprule
 & & T  &$a$  & $c/a$ & $B_T$  & $B_T^{\prime}$ & $B_T^{\prime \prime}$  \\
     & &  (K) & (a.u.) & & (kbar) &  & (kbar$^{-1}$) \\
\hline
This study &LDA & 0 & $5.133$ & $1.579$ & $4489$ & $4.7$  & $-0.0015$\\ 
This study &PBEsol & $0$ & $5.159$ & $1.579$ &$4328$ & $4.7$ & $-0.0016$ \\
This study &PBE & $0$ & $5.204$ & $1.578$ & $3984$ & $4.8$ & $-0.0018$ \\  
   & & $295$ & $5.212$ & $1.579$ & $3914$ & $4.9$ & \\
   & & $1500$ & $5.242$ & $1.585$ &  &  & \\
 \newline  \\
 Calc.~\cite{liu_structural_2011} & LDA & 0& $5.132$ & $1.574$& $4470$ & $4.63$ &\\ 
 \newline  \\
  Calc.~\cite{ma_electronic_2005} & PBE & 0& $5.20$ & $1.577$& $4080$& $3.6$ &\\
Calc.~\cite{deng_elastic_2009} & PBE & 0& $5.187$ & $1.577$ & $3960$& & \\
 \newline  \\

 Expt.~\cite{kenichi_bulk_2004} & & $300$& $5.1680$ & $1.5794$ & $3950$ & $4.5$ & \\
\botrule
\end{tabular}
\label{table:1}
\end{center}
% \end{minipage}
% \end{center}

\end{table*}

\begin{table*}
  \caption{\label{table:2} Elastic constants compared with experiment and previous calculations. 
We report also the bulk modulus ($B$), the Young's modulus ($E$), the shear modulus ($G$), and Poisson's ratio ($\nu$) of polycrystalline osmium derived from the elastic constants. PBE elastic constants including zero-point motion (ZPM) and adiabatic QHA elastic constants at 300 K are also reported. Frozen-ion elastic constants are obtained by applying a uniform strain to atomic positions, omitting further atomic relaxation.
}
\begin{ruledtabular}
\begin{tabular}{ccccccccccccc}
 & T &$a$ & $\frac{c}{a}$  & $C_{11}$  & $C_{12}$ & $C_{13}$ & $C_{33}$ & $C_{44}$ & $B$ & $E$ & $G$ & ${\nu}$ \\
  & (K) & (a.u.) &  & (kbar) & (kbar) & (kbar) & (kbar) & (kbar) & (kbar) & (kbar) & (kbar) & \\
\hline
 This study (LDA) & $0$ & $5.133$ & $1.579$ & $8191$ & $2529$  & $2487$ & $9161$ & $2851$ & $4501$ & $7173$ & $2906$ & $0.234$ \\
 Ref.~\cite{fan_potential_2006} (LDA) & 0 &  &  & $8087$ & $2647$  & $2437$ & $8886$ & $2712$ & $4454$& $6931$ &$2793$&$0.241$   \\
Ref.~\cite{minisini_elastic_2008} (LDA) & 0 &  &  & $7890$ & $3080$  & $2390$ & $9580$ & $2890$ & $4559$& $6928$ &$2778$&$0.247$   \\
Ref.~\cite{yu_calculations_2010} (LDA) & $0$ &  &  & $8390$ & $2460$  & $2570$ & $9250$ & $2790$ & $4576$& $7251$ &$2934$&$0.236$   \\
 Ref.~\cite{fast_elastic_1995} (LDA) & $0$ & $5.202$ & $1.578$ & $8945$ & $2492$  & $2456$ & $10164$ & $1622$ & $4755$& $6402$ &$2511$&$0.275$   \\
 \\
 This study (PBEsol) & $0$ & $5.156$ & $1.579$ & $7948$ & $2413$ & $2366$ & $8907$ & $2788$ & $4339$ & $6999$ & $2842$ & $0.231$  \\
 \\
    This study (PBE) & $0$ & $ $ & $ $ & $7332$ & $ 2202 $  & $2170$ & $8224 $ & $2585$ & $ 3992$ & $6476 $ & $2633 $ & $0.230 $ \\
  PBE (Frozen-ions) & $0$ & $ $ & $ $ & $7366$ & $2168$  & $2171$ & $8224$ & $ 2589$ & $ 3993$ & $6507$ & $2649$ & $0.228$ \\
 This study (PBE) & $0$+ZPM & $5.207$ & $1.578$ & $7262$ & $2188$  & $2162$ & $8135$ & $2555$ & $3960$ & $6405$ & $2603$ & $0.230
 $ \\
 This study (PBE) & $301$ & $5.212$ &  $1.578$ & $7104$ & $2194$  & $2174$ & $7948$ & $2465$ & $3911$ & $6213$ & $2515$ & $0.235$ \\
Ref.~\cite{fan_potential_2006} (PBE) & 0 &  &  & $7301$ & $2469$  & $2305$ & $7983$ & $2469$ & $4081$& $6231$ &$2501$&$0.246$   \\
 Ref.~\cite{minisini_elastic_2008} (PBE) & 0 &  &  & $7150$ & $2740$  & $2020$ & $8700$ & $2650$ & $4058$& $6340$ &$2557$&$0.240$   \\
  Ref.~\cite{deng_elastic_2009} (PBE) & 0 &  &  & $7480$ & $2090$  & $2070$ & $8220$ & $2610$ & $3957$& $6618$ &$2710$&$0.221$   \\
  \\
Expt.~\cite{panteaElasticConstantsOsmium2009c}& 0 &  &  & $7633$ & $2279$  & $2180$ & $8432$ & $2693$ & $4105$ & $6737$& $2747$ & $0.226$ \\
\end{tabular}
\end{ruledtabular}
\label{tab:elastic}
\end{table*}

\begin{figure}
\centering
\includegraphics[width=\linewidth]{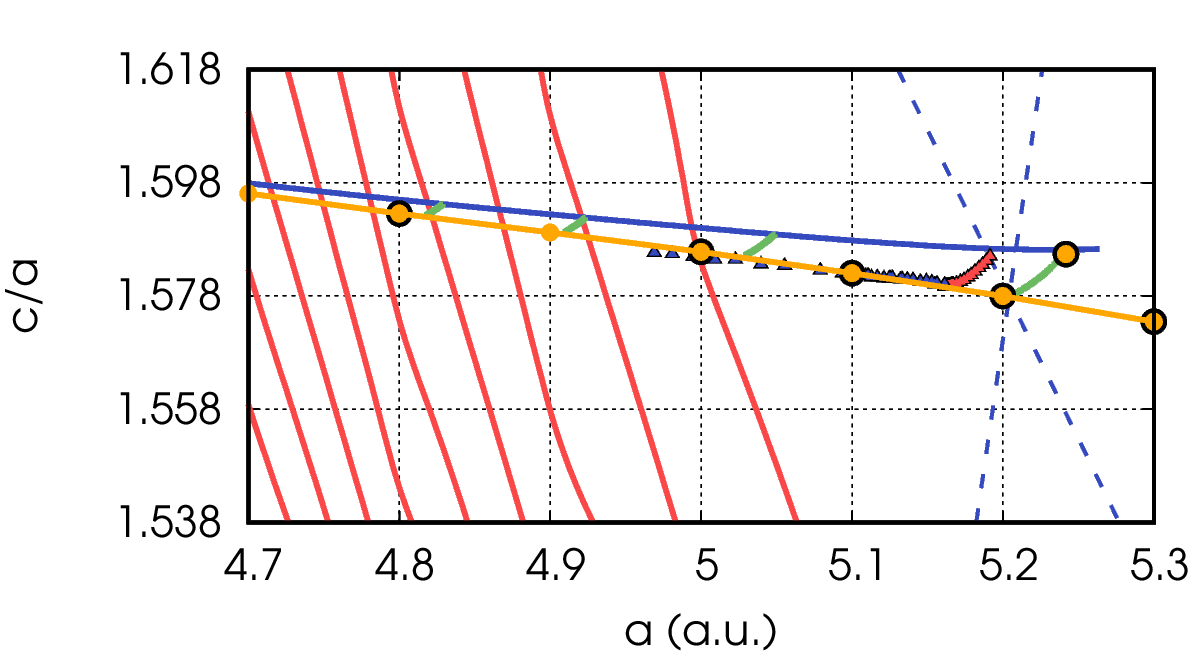}
\caption{Energy contours in the 
$a$ and $c/a$ crystal parameter plane (red lines). The intersection of the dashed blue lines pinpoints the energy minimum. The yellow and blue curves depict the curve where the stress is a uniform pressure at $0$ K and $1500$ K, respectively. The yellow dots mark the seven distinct geometries used to map the stress-pressure curve at $0$ K. Black circles identify the geometries where TDECs were computed using the QHA. Green lines illustrate isobars at $0$, $500$, $1000$, and $1500$ kbar. The specific geometry at $1500$ K and $0$ kbar is also highlighted with a yellow dot. The thin dotted lines represent the $7\times 5$ grid of crystal parameters used for free energy calculations. Experimental data is overlaid: blue triangles show room-temperature stress-pressure curve from Ref.~\cite{kenichi_bulk_2004}, and red triangles show $0$ kbar isobar data from $0$ to $1300$ K, as recommended in Ref.~\cite{arblaster_crystallographic_2013}.}
\label{fig:energy_os}
\end{figure}

\begin{figure}
\centering
\includegraphics[width=\linewidth]{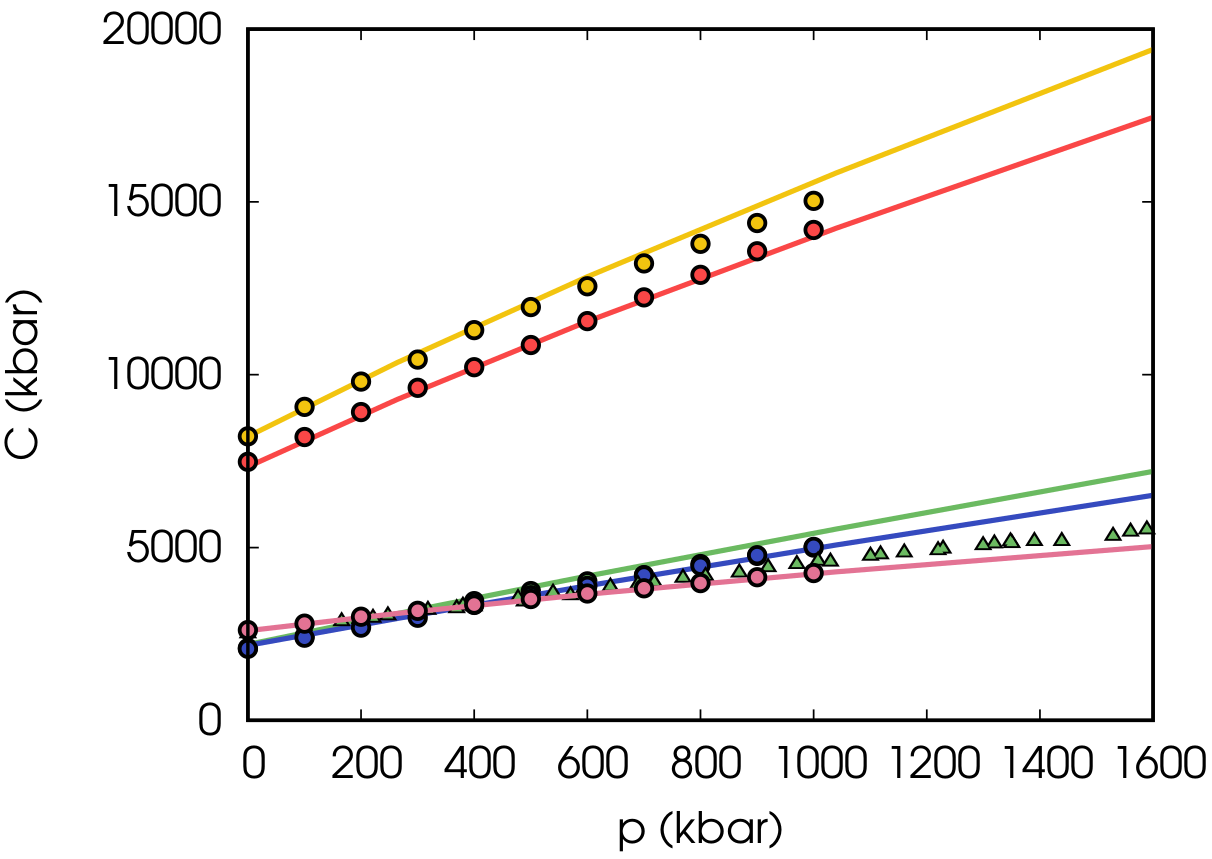}
\caption{Pressure-dependent ECs of osmium (PBE calculation): 
$C_{11}$ (red line), $C_{33}$ (yellow line), $C_{44}$ (pink line), $C_{12}$ (green line), $C_{13}$ (blue line). Dots represent PBE calculations from Ref.~\cite{deng_elastic_2009} (same color convention). Green triangles show the experimental measurements of $C_{44}$ from Ref.~\cite{jingyi_liu_high-pressure_2022}.}
\label{fig:el_cons_p}
\end{figure}

\begin{figure}
\centering
\includegraphics[width=\linewidth]{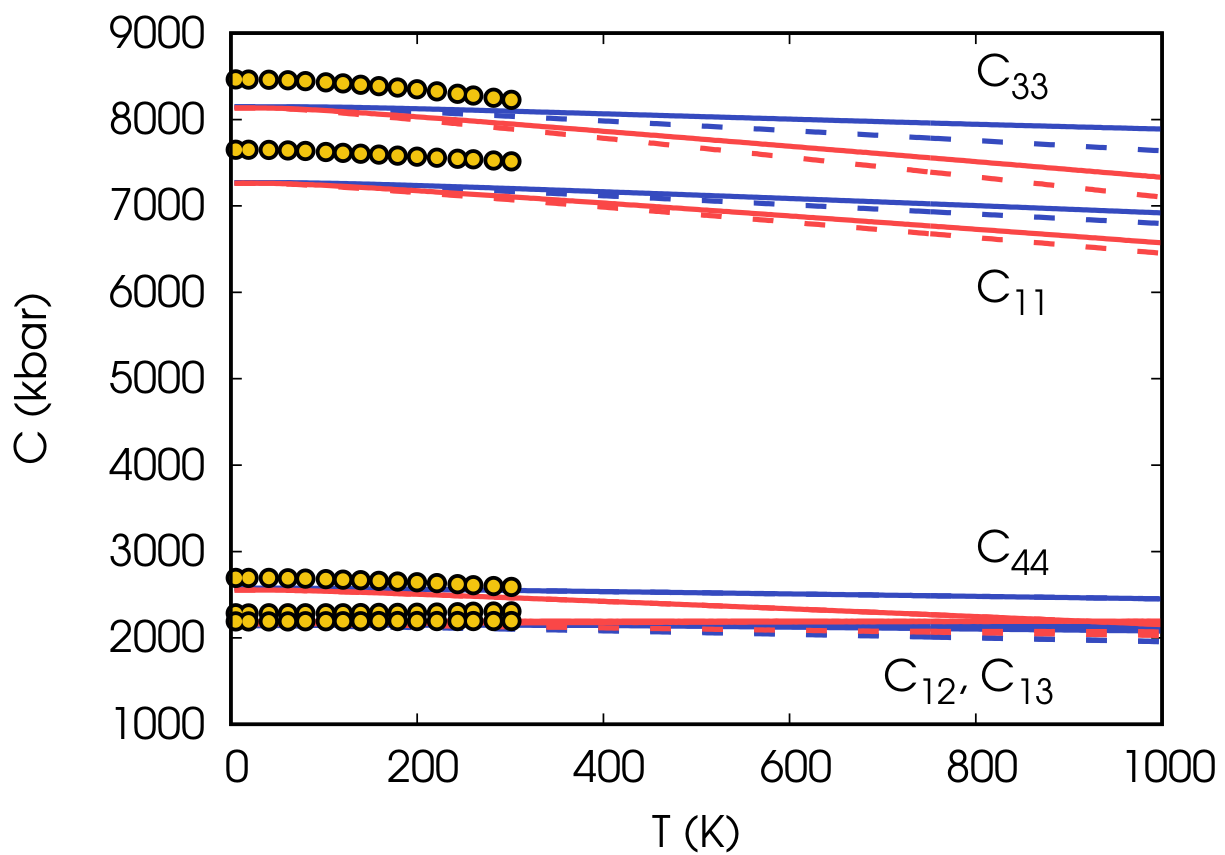}
\caption{TDECs of osmium (PBE calculation). Dashed lines: isothermal; solid lines: adiabatic. Blue curves: QSA; red curves: QHA. Yellow circles: experimental data from Ref.~\cite{panteaElasticConstantsOsmium2009c}.
}
\label{fig:elastic_os}
\end{figure}

\section{Theory: Thermodynamics and elastic constants}

All thermodynamic calculations employ our in-house software package \texttt{thermo\_pw}~\cite{dal_corso_thermo_pw_2022}, which implements the quasi-harmonic approximation (QHA) as detailed in previous works~\cite{dal_corso_elastic_2016,palumbo_lattice_2017,pal2,malica_temperature-dependent_2019,malica_quasi-harmonic_2020,malica_temperature_2020,malica_quasi-harmonic_2021}. The workflow for calculating TDECs in hcp systems was established in Ref.~\cite{gong_high-temperature_2024}. For completeness, we outline the key theoretical framework below.

Within the QHA formalism, the Helmholtz free energy of a hexagonal crystal depends on temperature $T$ and lattice parameters 
$\xi=(a,c/a)$, decomposing into three contributions:
\begin{equation}
F(\xi,T)= U(\xi) + F_{ph}(\xi,T) + F_{el}(\xi,T),
\label{eq:free_ener}
\end{equation}
where $U(\xi)$ represents the static DFT energy, 
$F_{ph}(\xi,T)$ the phonon vibrational free energy, and $F_{el}(\xi,T)$ the electronic excitation term. The vibrational component derives from phonon frequencies ${\omega}_{\eta}({\bf q},\xi)$:
\begin{eqnarray} \label{equ4}
F_{vib}(\xi, T) & =& \frac{1}{2N} \sum_{\mathbf q \eta} \hbar \omega_{\eta} \left(\mathbf q,
\xi \right) \nonumber\\
& +& {\frac{1}{N \beta}} \sum_{\mathbf q \eta} \ln \left[1 - \exp \left(- \beta \hbar \omega_{\eta}(\mathbf q, \xi)\right) \right].
\end{eqnarray} 
Here $\hbar$ is the reduced Planck's constant,
$N$ is the number of cells of the solid (equal to the number of ${\bf q}$ points), $\beta={\frac{1}{k_B T}}$ where $k_B$ is Boltzmann's constant, ${\bf q}$ denotes phonon wavevectors, and 
$\eta$ indexes vibrational modes. The electronic excitation term $F_{el}(\xi_i,T)$ follows from the density of states within the rigid-band approximation~\cite{malica_quasi-harmonic_2021}.
In these expressions, phonon frequencies are calculated with an electronic temperature of $0$ K, and the energy $U(\xi)$ is temperature-independent. This approach assumes that electronic excitations have a minimal impact on the phonon free energy.
Alternatively, electronic excitation effects can be included in $U(\xi)$ and in the phonon frequencies using Fermi-Dirac occupations at a specific temperature. 
However, this procedure is numerically more demanding and necessitates modifications to the free energy expression for full consistency between free energy and entropy, as discussed in Ref.~\cite{allen_anharmonic_2015}. 
In this paper, we test this procedure at $1500$ K, but for all other calculations, we use phonon frequencies obtained with an electronic temperature of $0$ K.

The equilibrium state under stress tensor 
$\boldsymbol{\sigma}$ emerges from minimizing the functional:
\begin{equation}
G_{\boldsymbol{\sigma}}(\xi,T)=F(\xi,T)+V\sum_{j=1}^6 \sigma_j \varepsilon_j(\xi),
\label{gibbs}
\end{equation}
where $V$ is the volume of one unit cell.
We have:
\begin{equation}
\sigma_j=-{1\over V} {\frac{\partial F(\xi,T)}{\partial \varepsilon_j}}.
\label{eq:p}
\end{equation}
Hence the crystal parameters that minimizes $G_{\boldsymbol{\sigma}}(\xi,T)$ are the crystal parameters that correspond to stress $\boldsymbol{\sigma}$.
Using for the stress a uniform pressure $p$, we find the crystal parameters corresponding to each pressure at any temperature ($\xi_p$). 
 
The hexagonal thermal expansion tensor contains two independent components:
\begin{eqnarray}
\alpha_{xx} &=& \alpha_{yy} = \frac{1}{a} \frac{da}{dT}, \\
\alpha_{zz} &=& \frac{1}{c} \frac{dc}{dT},
\end{eqnarray}
with volume expansion $\beta=2\alpha_{xx}+\alpha_{zz}$.

Isothermal ECs derive from second derivatives of $F(\varepsilon_{i},T)$:
\begin{equation}
\tilde{C}^T_{ij} = \frac{1}{V} \left.\frac{\partial^2 F}{\partial \varepsilon_{i} \partial \varepsilon_{j}}\right|_T,
\label{tdec}
\end{equation}
evaluated for five distinct strain configurations: $(\varepsilon,0,0,0,0,0)$,
$(0,0,\varepsilon,0,0,0)$, $(\varepsilon,0,\varepsilon,0,0,0)$,
$(\varepsilon,\varepsilon,0,0,0,0)$, and $(0,0,0,\varepsilon,0,0)$. In these cases 
${1\over V} {\partial^2 F \over \partial \varepsilon^2}$
gives the following combinations of ECs 
$\tilde C_{11}$, $\tilde C_{33}$, $\tilde C_{11} +\tilde C_{33} + 2 \tilde C_{13}$, $2 \tilde C_{11}  + 2 \tilde C_{12}$, and $\tilde C_{44}$
respectively. 
When the equilibrium reference configuration has a non vanishing stress $\sigma_{ij}$, to obtain the stress-strain ECs we apply the correction~\cite{barron_second-order_1965} (in cartesian notation):
\begin{eqnarray}
C^T_{ijkl} =  \tilde C^T_{ijkl} &-&
{1\over 2} \Big(2 \sigma_{ij} \delta_{kl}
-{1\over 2} \sigma_{ik} \delta_{jl}
-{1\over 2} \sigma_{il} \delta_{jk} \nonumber \\
&-&{1\over 2} \sigma_{jk} \delta_{il}
-{1\over 2} \sigma_{jl} \delta_{ik} \Big),
\label{eqsd2}
\end{eqnarray}
which simplifies for hydrostatic pressure $\sigma_{ij}=-p\delta_{ij}$ to:
\begin{equation}
C^T_{ijkl} = \tilde C^T_{ijkl} + {p \over 2} \left(2 \delta_{i,j} \delta_{k,l}
- \delta_{i,l} \delta_{j,k} - \delta_{i,k} \delta_{j,l}  \right).
\label{eq:correct_p}
\end{equation}
The second derivatives of the free energy are calculated as described in 
Ref.~\cite{dal_corso_elastic_2016} taking as equilibrium configuration a subset of parameters $\xi_i$ along the stress-pressure curve at $0$ K. The values of $\xi_i$ along this curve are given in the supplemental material,~\cite{supplemental}
together with the pressure present in each configuration.
The ECs at any other set of parameters $\xi_p$ at temperature $T$ and pressure $p$ are obtained by interpolation by a fourth-degree polynomial.

Adiabatic ECs follow from the transformation:
\begin{equation}
C^S_{ijkl} = C^T_{ijkl} + \frac{TV b_{ij} b_{kl}}{C_V},
\end{equation}
where $b_{ij}$ are the thermal stresses:
\begin{equation}
b_{ij} = - \sum_{kl} C^T_{ijkl} \alpha_{kl}.
\end{equation}
\par

\section{Computational details}

First-principles calculations are carried out within density functional theory (DFT) using the Quantum ESPRESSO package~\cite{qe1, qe2}. Three exchange-correlation functionals are employed: the LDA,~\cite{lda} the PBEsol,~\cite{pbesol} and the PBE generalized gradient approximations.~\cite{pbe} While structural properties and zero-temperature ECs are computed with all three functionals, TDECs and thermodynamic properties are evaluated exclusively with PBE.

Electron-ion interactions are treated with projector augmented wave (PAW) pseudopotentials~\cite{corso_dalcorsopslibrary_2022} including $5s$ and $5p$ semicore states together with the $5d$, $6s$ valence states (16 electrons per atom), with nonlinear core corrections applied.~\cite{louie_nonlinear_1982} The pseudopotentials used are \texttt{Os.pz-spn-kjpaw\_psl.1.0.0.UPF} (LDA),
\texttt{Os.pbesol-spn-kjpaw\_psl.1.0.0.UPF} (PBEsol) and
\texttt{Os.pbe-spn-kjpaw\_psl.1.0.0.UPF} (PBE).
We expand the wavefunctions in plane waves with a cutoffs of 80 Ry (wavefunctions) and 320 Ry (charge density).
The Brillouin zone integrations are made using the Methfessel-Paxton smearing ($\sigma$ = 0.02 Ry)~\cite{mp} with a $48 \times 48 \times 32$ ${\bf k}$-point mesh (See supplemental material~\cite{supplemental} for a test of the convergence of elastic constants with {\bf k} points at $0$ K).
Calculations are performed on the Leonardo supercomputer at CINECA using a GPU optimized version of \texttt{thermo\_pw}.~\cite{gong_alternative_2025}

The equilibrium lattice parameters are determined by minimizing the total energy on a $7 \times 5$ grid of ($a$, $c/a$) values, covering pressures from $-180$ kbar to $2300$ kbar. Phonon dispersions and Helmholtz free energies are computed in all these grid points and at seven selected points along the stress-pressure curve (Table~\ref{table:1} of the supplemental material~\cite{supplemental}). Among these, five geometries (i = $2$, $4$, $5$, $6$, $7$) are used to evaluate the QHA TDECs via second derivatives of the free energy with respect to strain (geometry $1$ is at the highest pressure). For each geometry, five strain types are applied: base-centered orthorhombic (types 1 and 3), hexagonal (types 2 and 4) and monoclinic (type 5).
Each strain is sampled at six values ($\epsilon \in [-0.0125, 0.0125]$, $\Delta\epsilon = 0.005$), requiring phonon and electronic density of states (DOS) calculations for $150$ configurations.

Phonon frequencies are obtained via density functional perturbation theory (DFPT)~\cite{rmp, dfptPAW} on a $6 \times 6 \times 6$ {\bf q}-point grid, followed by Fourier interpolation onto a $200 \times 200 \times 200$ mesh for thermodynamic integration. The electronic DOS is computed on a $100 \times 100 \times 100$ {\bf k}-point grid.

\begin{figure}
\centering
\includegraphics[width=\linewidth]{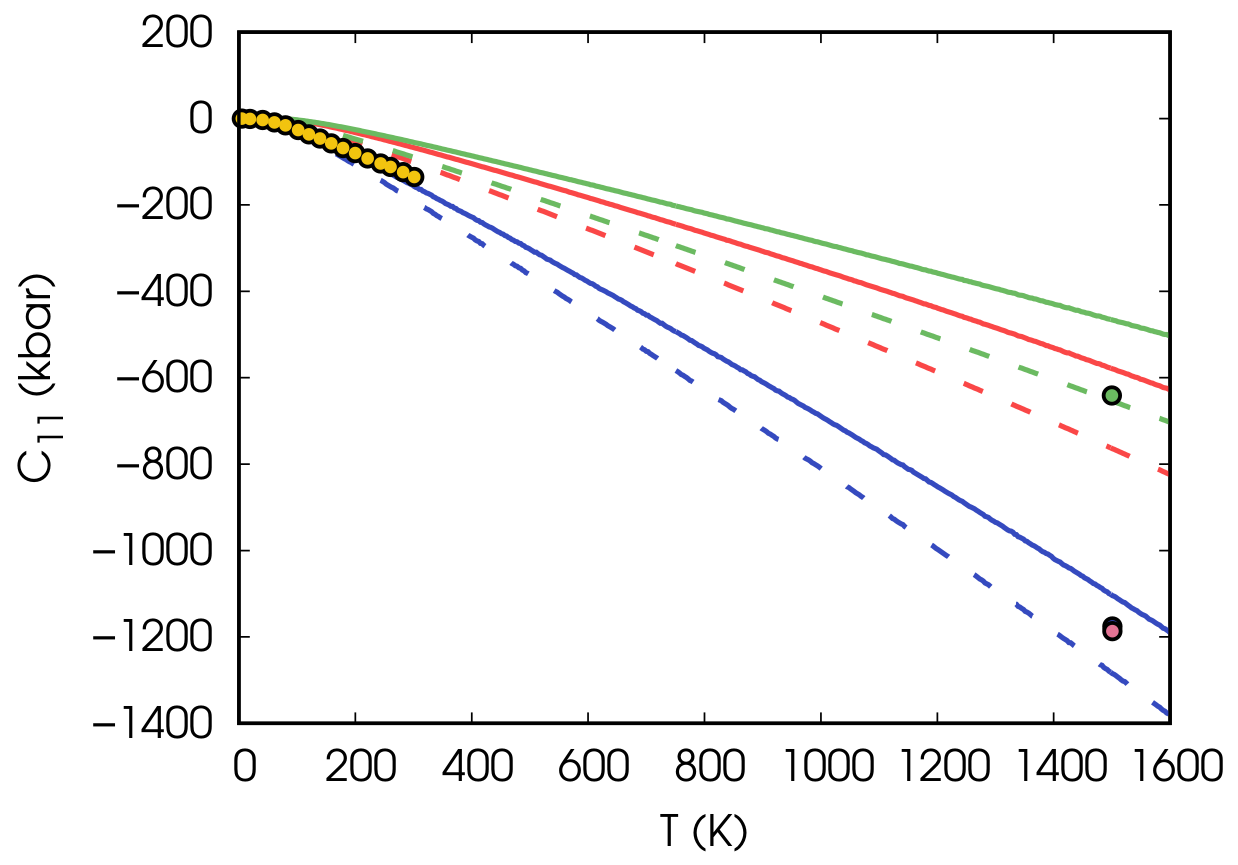}
\caption{Temperature dependence of the 
$C_{11}$ EC of osmium (PBE calculation, with $0$ K value subtracted). Dashed lines: isothermal; solid lines: adiabatic. Red (green) curves: QSA using V-ZSISA (full grid interpolation). Blue curves: QHA using V-ZSISA. Yellow circles: experimental data from Ref.~\cite{panteaElasticConstantsOsmium2009c}. Blue (green) dots at $1500$ K: isothermal QHA (QSA) results at the $1500$ K geometry. Pink dots at $1500$ K: isothermal QHA obtained with Fermi-Dirac occupations.}
\label{fig:c11_os}
\end{figure}

\begin{figure}
\centering
\includegraphics[width=\linewidth]{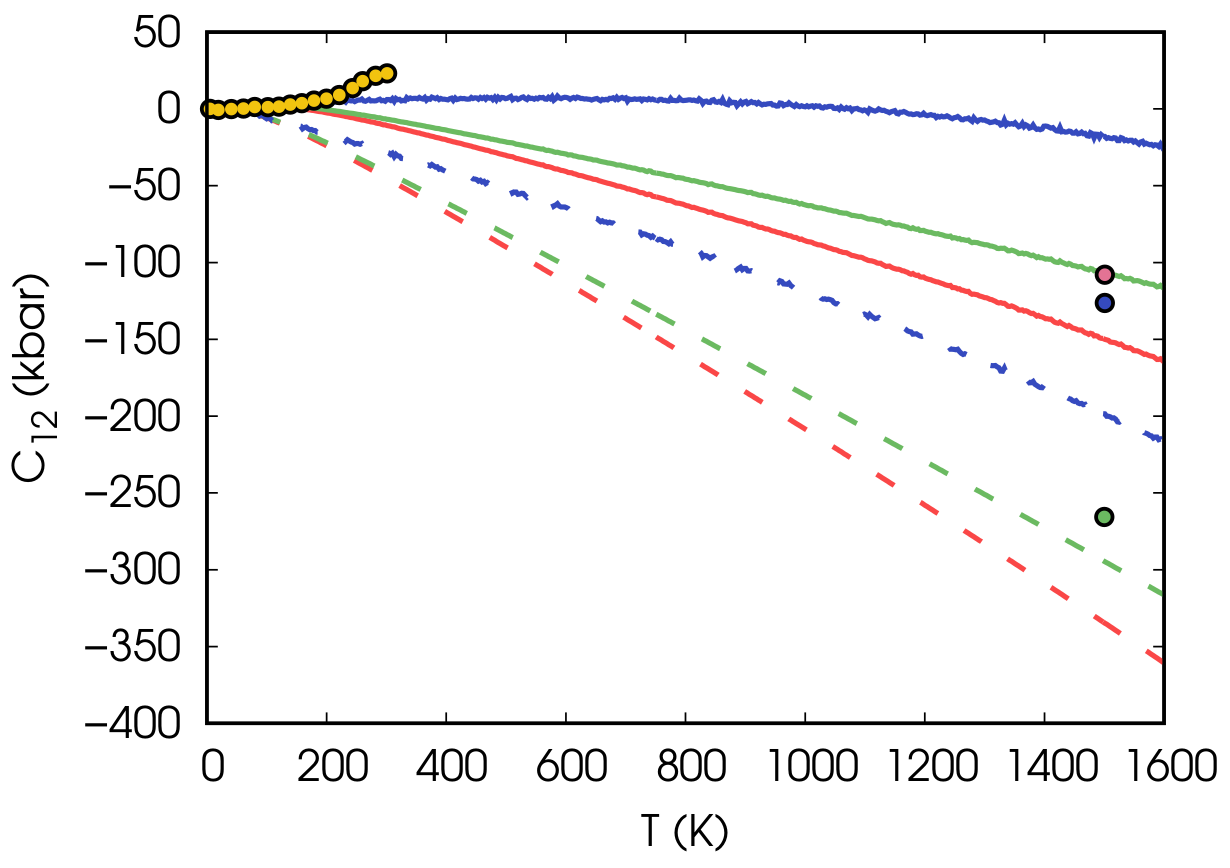}
\caption{Temperature dependence of the 
$C_{12}$ EC of osmium (PBE calculation, with $0$ K value subtracted). Line and dot conventions are identical to Fig.~\ref{fig:c11_os}.}
\label{fig:c12_os}
\end{figure}

\begin{figure}
\centering
\includegraphics[width=\linewidth]{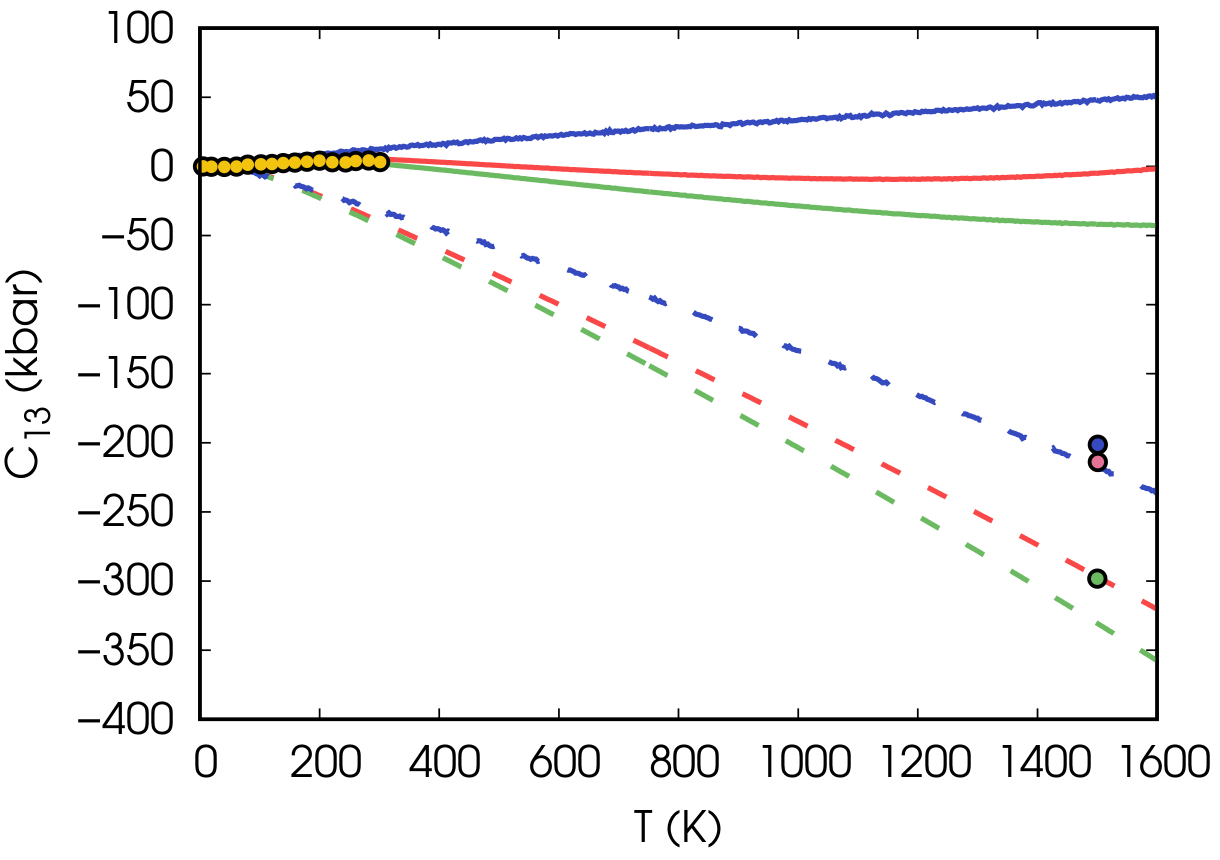}
\caption{Temperature dependence of the 
$C_{13}$ EC of osmium (PBE calculation, with $0$ K value subtracted). Line and dot conventions are identical to Fig.~\ref{fig:c11_os}.}
\label{fig:c13_os}
\end{figure}

\begin{figure}
\centering
\includegraphics[width=\linewidth]{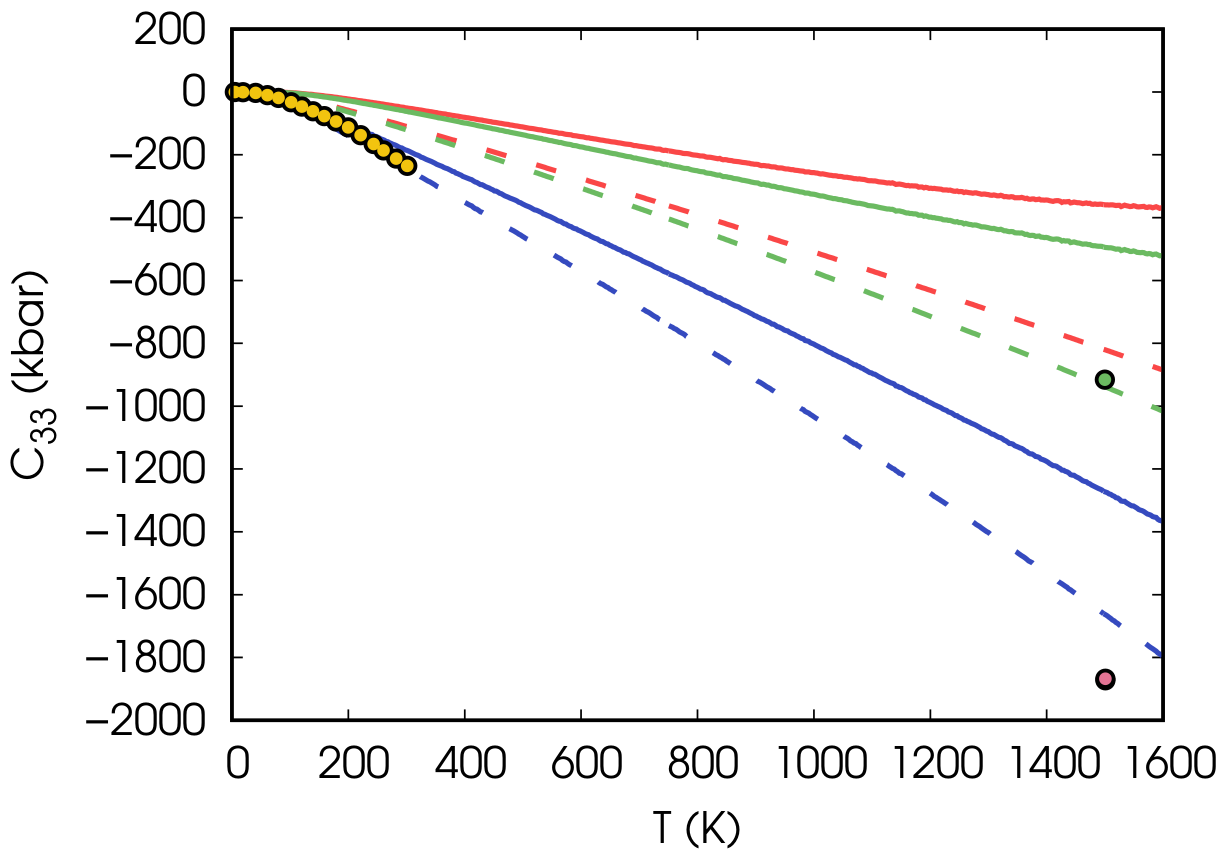}
\caption{Temperature dependence of the 
$C_{33}$ EC of osmium (PBE calculation, with $0$ K value subtracted). Line and dot conventions are identical to Fig.~\ref{fig:c11_os}.}
\label{fig:c33_os}
\end{figure}

\begin{figure}
\centering
\includegraphics[width=\linewidth]{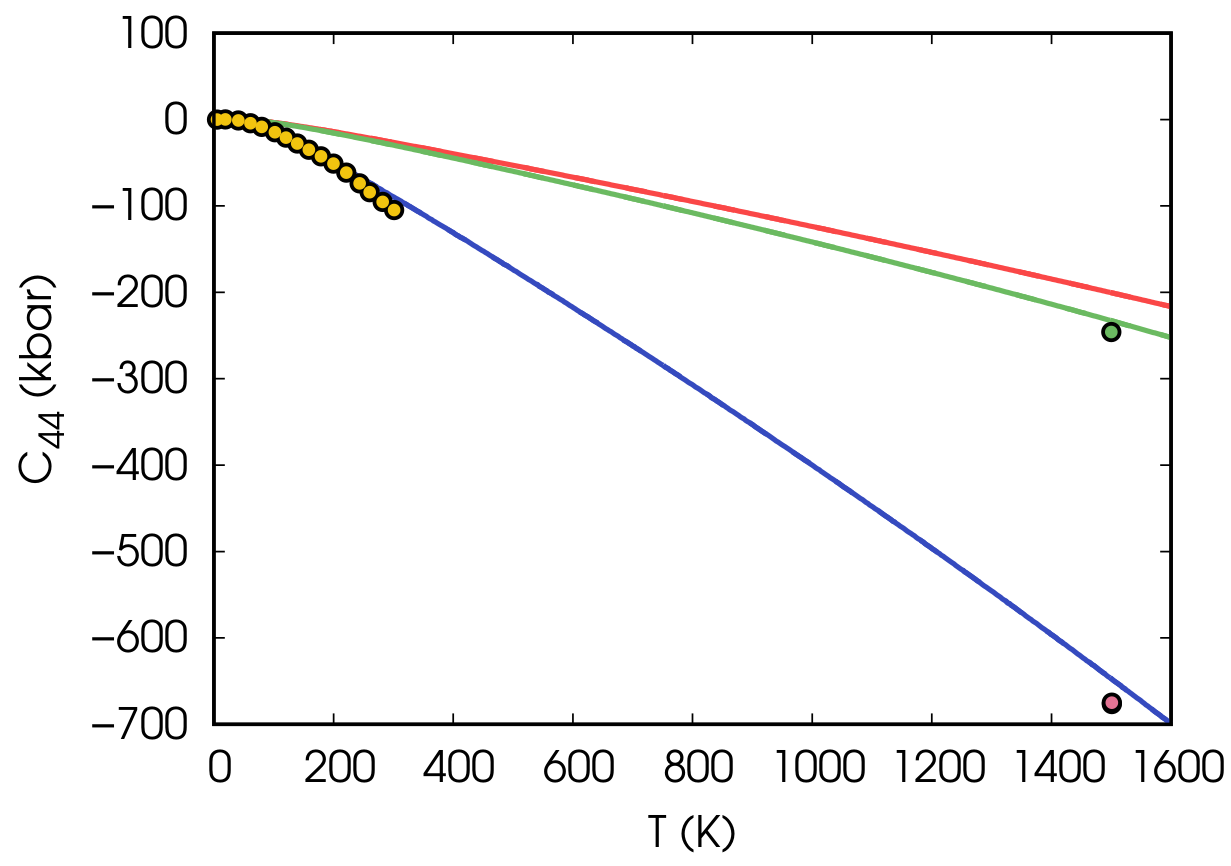}
\caption{Temperature dependence of the 
$C_{44}$ EC of osmium (PBE calculation, with $0$ K value subtracted). Line and dot conventions are identical to Fig.~\ref{fig:c11_os}.}
\label{fig:c44_os}
\end{figure}

\section{Results and Discussion}

We present in Tab.~\ref{table:1} the crystal parameters,
the bulk modulus and the pressure derivative of the bulk modulus obtained from a Birch-Murnaghan fit of the total energy with respect to the volume along the stress-pressure curve. 
Assuming an experimental value of $a=5.160$ a.u. obtained by extrapolating at $0$ K the
$300$ K value of Ref.~\cite{kenichi_bulk_2004}, the LDA, PBEsol and PBE errors are $-0.5 \%$,
$<0.1 \%$ and $0.8 \%$ respectively, while the value of $c/a$ is essentially accurate with all functionals.
Our values are in agreement with the finding of Ref.~\cite{palumbo_lattice_2017} and we refer to that paper for a comparison of these values with those found in the literature.

In Tab.~\ref{table:2} we present the calculated ECs at $0$ kbar and $0$ K and compare the performance of the three functionals with experiments and with previous calculations. In the same table we present also the bulk modulus, Young modulus, shear modulus and Poisson ratio of polycrystalline osmium obtained with the calculated ECs and the Voigt-Reuss-Hill approximation.~\cite{newnham_properties_2004}
Taking the values given in Ref.~\cite{panteaElasticConstantsOsmium2009c} at $0$ K as experimental values, the errors of $C_{11}$, $C_{12}$, $C_{13}$, $C_{33}$, $C_{44}$ are $558$ kbar ($7 \%$),
$250$ kbar ($11 \%$), $307$ kbar ($14 \%$), $729$ kbar 
($9 \%$), and $158$ kbar ($6 \%$) for LDA, they decrease to $315$ kbar ($4 \%$), $134$ kbar ($6 \%$), $186$ kbar ($9 \%$), $475$ kbar ($6 \%$), and $95$ kbar ($4 \%$) for PBEsol, and further decrease to
$-301$ kbar ($-4 \%$), $-77$ kbar ($-3 \%$), $-10$ kbar ($-0.5 \%$), $-208$ kbar ($-2 \%$), and $-108$ kbar ($-4 \%$) for PBE.
We note also that zero point motion (ZPM) decreases further the calculated ECs even for an heavy atom as osmium. We can estimate its effects from the free energy of the distorted structures in the PBE case. The decrease for $C_{11}$, $C_{12}$,
$C_{13}$, $C_{33}$, and $C_{44}$ is $70$ kbar ($0.9 \%$), $14$ kbar ($0.6 \%$), $8$ kbar ($0.4 \%$), $89$ kbar ($1 \%$), and $30$ kbar ($1 \%$). When accounting for the ZPM, the PBE functional yields $C_{12}$, $C_{13}$, and $C_{33}$ that more closely align with experimental values, while PBE and PBEsol becomes comparable for $C_{11}$ and $C_{44}$. As demonstrated in our previous work, the functional selection primarily affects the ECs at $0$ K, while temperature dependencies remain consistent across different functionals. Consequently, we calculated the TDECs solely using the PBE functional.
Comparing the bulk moduli derived from ECs with those obtained from the equation of state, we find a reasonable agreement (the errors are $12$ kbar, $11$ kbar, and $8$ kbar for LDA, PBEsol, and PBE, respectively).

In Fig.~\ref{fig:energy_os} we show the contour plots of the PBE energy, together with the stress-pressure curves at $0$ K and at $1500$ K. At $0$ K there is a very good agreement between our data and those measured at room temperature in Ref.~\cite{kenichi_bulk_2004}. Note however a small shift of the pressure since with PBE the value of $a$ at the minimum is $0.8 \%$ larger than the experiment. At variance with what found in beryllium,~\cite{gong_high-temperature_2024}
where the $0$ kbar isobar is almost parallel to the stress-pressure curve at $0$ K, in osmium it is almost perpendicular. Osmium is therefore a more strict test of the V-ZSISA approximation.
Moreover, both $a$ and $c/a$ increase with temperature while along the stress-pressure curve when $a$ increases $c/a$ decreases. Therefore the stress-pressure curve at $1500$ K is above the $0$ K one while it was below in beryllium. The predicted behavior of osmium's crystal parameters with temperature and pressure is in agreement with what found in a previous paper~\cite{palumbo_lattice_2017}
and in experiment.~\cite{arblaster_crystallographic_2013} 

In Fig.~\ref{fig:el_cons_p}, we plot, at $0$ K, the pressure dependent ECs in the range $0$ kbar-$1600$ kbar and compare
them with the previous theoretical PBE calculation~\cite{deng_elastic_2009} and with the recent experimental data on $C_{44}$~\cite{jingyi_liu_high-pressure_2022}.
There is a quite good agreement between our data and those of Ref.~\cite{deng_elastic_2009} in all the range in which they have been calculated. The agreement of $C_{44}$ with experiment is also quite good.

We started the TDECs calculation by testing the ZSISA approximation as described in Ref.~\cite{gong_high-temperature_2024}. As shown in Tab.~\ref{table:2} the effect of relaxations on the ECs of osmium is quite small. Comparing the relaxed ions and the frozen ions calculations, we see that the change is 
$\Delta C_{11}=-\Delta C_{12}=34$ kbar 
($0.4 \%$ for $C_{11}$ and $1.5 \%$ for $C_{12}$).
We therefore do not expect a big effect of the
ZSISA approximation. We show in the supplemental material~\cite{supplemental} that indeed the effect is quite small. Maximum differences at $1500$ K are of about $10$ kbar.

QSA and QHA ECs both calculated within the V-ZSISA approximation are shown in Fig.~\ref{fig:elastic_os}.
$C_{11}$, $C_{33}$ and $C_{44}$ decrease more rapidly within QHA than within QSA. However, 
on this extended scale the experimental variation of the ECs with temperature is small and it is difficult to compare the two approximations. Both of them are in reasonable agreement with experiment.
On this scale, the QSA ECs, calculated via V-ZSISA along the stress-pressure curve or beyond V-ZSISA by grid interpolation along the $0$ kbar isobar, show negligible differences when the appropriate volume $V(T)$ is used for V-ZSISA interpolation at each temperature. A comparison of these approaches on this scale is included in the supplemental material,~\cite{supplemental} while a detailed illustration is in the following figures.

 \begin{figure}
\centering
\includegraphics[width=\linewidth]{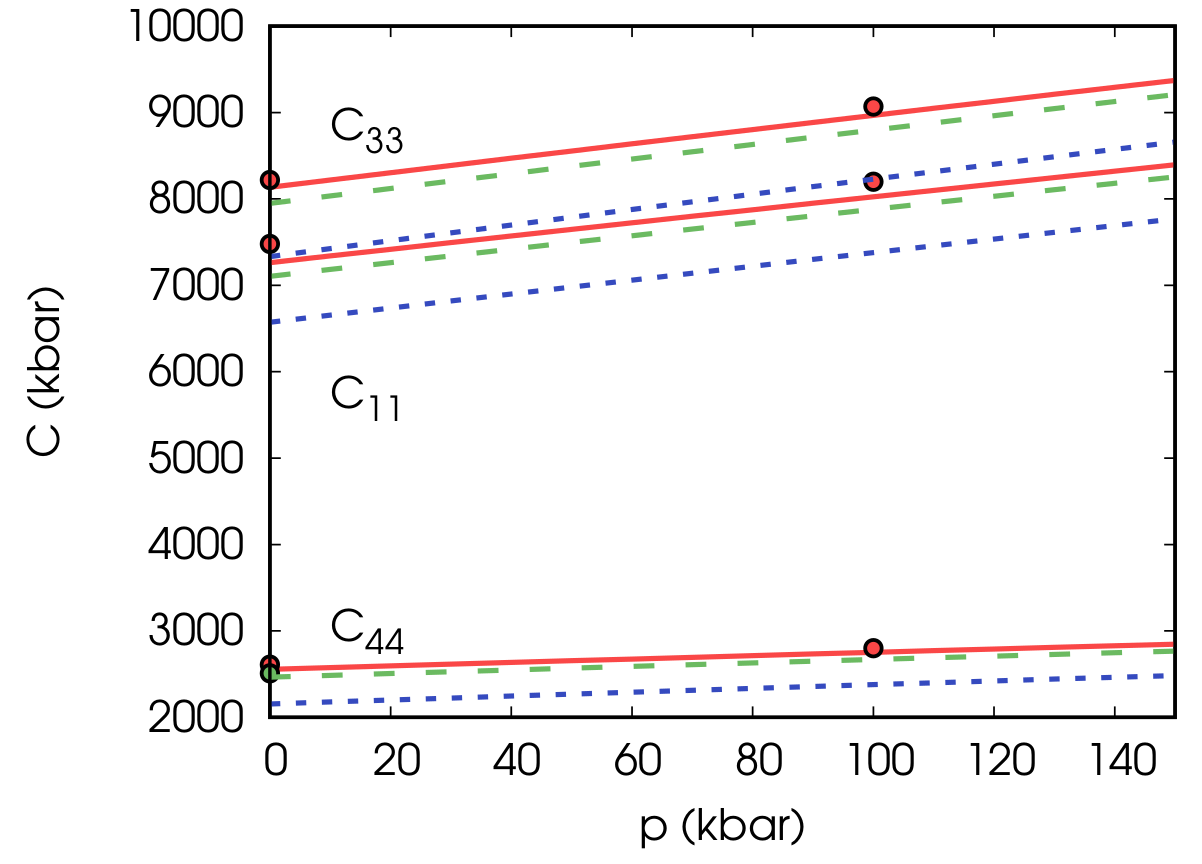}
\caption{PBE QHA pressure dependence of osmium adiabatic ECs 
$C_{11}$, $C_{33}$, and $C_{44}$ at $4$ K (red lines), $301$ K (green dashed lines), $1000$ K (blue dotted lines), and $1501$ K (orange dot-dashed lines). Dots represent $0$ K PBE calculations from Ref.~\cite{deng_elastic_2009}.}   
\label{fig:ec_pressure}
\end{figure}

\begin{figure}
\centering
\includegraphics[width=\linewidth]{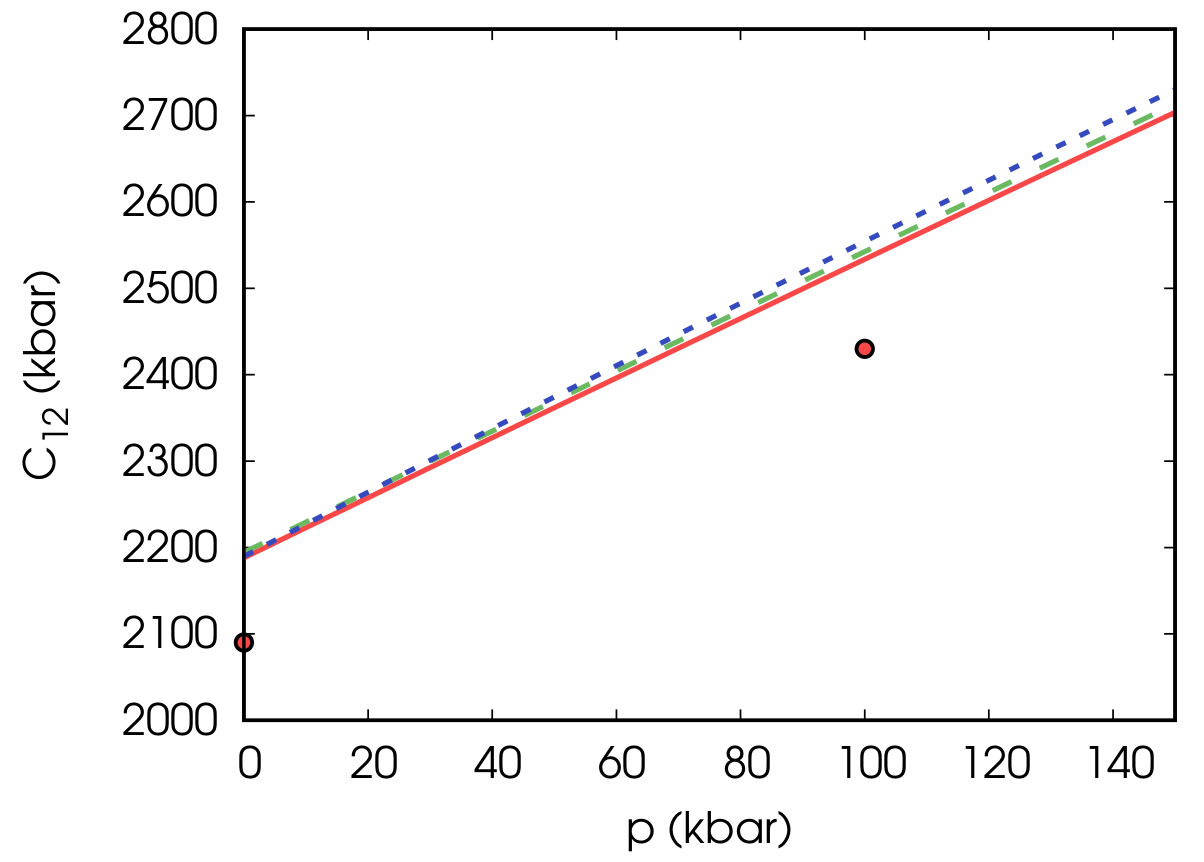}
\caption{PBE QHA pressure dependence of osmium adiabatic $C_{12}$ EC. Color and line style conventions are consistent with Fig.~\ref{fig:ec_pressure}.}
\label{fig:c12_pressure}
\end{figure}

\begin{figure}
\centering
\includegraphics[width=\linewidth]{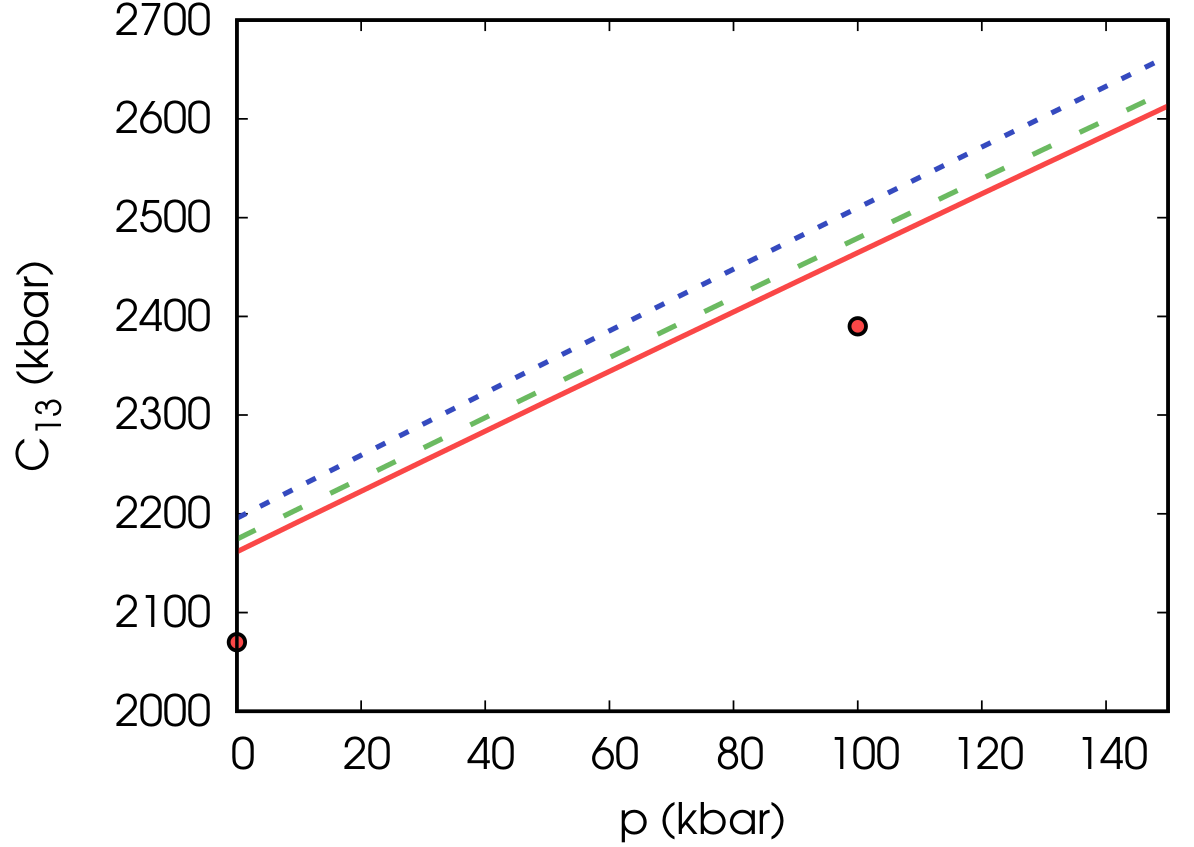}
\caption{PBE QHA pressure dependence of osmium adiabatic $C_{13}$ EC. Color and line style conventions are consistent with Fig.~\ref{fig:ec_pressure}.}
\label{fig:c13_pressure}
\end{figure}

The following five figures (from Fig.~\ref{fig:c11_os} to
Fig.~\ref{fig:c44_os}) present, in an enlarged scale, the TDECs obtained through various approximations, alongside experimental data. To facilitate comparison of temperature variation, the 
$0$ K values have been subtracted (see the supplemental material~\cite{supplemental} for the unshifted curves). Both isothermal (dashed lines) and adiabatic (solid lines) ECs are depicted.
Each figure incorporates three datasets. Firstly, the QHA ECs, derived within V-ZSISA, are represented in blue. The solid blue curve can be compared with experiments. Secondly, QSA ECs are shown. The latter are calculated within V-ZSISA, along the $0$ K stress-pressure curve (red curves), and
relaxing the V-ZSISA constraint, computing the
ECs along the $0$ kbar isobaric line (green curves).
The difference between the red and green curves quantifies the effect of the V-ZSISA approximation. It is reasonable to suppose that a comparable correction should be applied to the QHA ECs, if they were computed along the 0 kbar isobar. In general the effect of V-ZSISA is not very large, although well visible on the scale of these figures. 

As a singular validation test, we examined the geometry corresponding to $1500$ K, as detailed in Table~\ref{table:1}. Utilizing this geometry as the equilibrium configuration, we computed the phonon frequencies for all distorted configurations and subsequently determined the second derivatives of the free energy. This methodology eliminates the need for interpolation to obtain the QHA ECs at $1500$ K. The isothermal ECs derived from this approach are represented by the blue data points. Conversely, the values obtained at $4$ K are the isothermal QSA at $1500$ K and are indicated by green data points.

In general, across all figures, the green data point closely aligns with the isothermal QSA ECs calculated across the entire grid without the V-ZSISA approximation. The blue data point, however, exhibits a slight displacement from the V-ZSISA QHA ECs. This discrepancy is comparable to the difference observed between QSA ECs computed with V-ZSISA and those calculated across the full grid.

In the supplemental material,~\cite{supplemental} we also quantify the effect of electronic excitations on the TDECs by comparing calculations performed both with and without the electronic excitation term in the free energy. This effect proves to be quite small. We anticipate an even smaller impact when including electronic excitation effects on the phonon frequencies. As a single test, we repeated the TDEC calculation at the $1500$ K geometry using Fermi-Dirac occupations at this temperature in both the DFT and phonon frequency calculations. The results are indicated by pink (QHA at $1500$ K) dots in Figs.~\ref{fig:c11_os} to \ref{fig:c44_os}. As expected, the differences are minor, demonstrating that a more detailed treatment of electronic excitations is unnecessary for osmium within the investigated temperature range.

Examining each figure and comparing the ECs with experiment we can see that, with the exception of $C_{12}$ whose increase with temperature is slightly stronger than our theoretical prediction, the temperature dependence of all the other ECs is quite well predicted by our QHA calculation. 
From $5$ K to $301$ K the experimental decrease~\cite{panteaElasticConstantsOsmium2009c} is 
$136$ kbar ($1.8 \%$), $-23$ kbar ($-1 \%$),
$-3$ kbar ($-0.1$ \%), $236$ kbar ($2.8 \%$), and
$104$ kbar ($3.8 \%$) for $C_{11}$, $C_{12}$, 
$C_{13}$, $C_{33}$, $C_{44}$ respectively.
We can compare these decreases with those
predicted by QHA (within V-ZSISA) for the same
ECs:
$158$ kbar ($2.1 \%$), $-6$ kbar ($-0.3 \%$),
$-12$ kbar ($-0.6 \%$), $187$ kbar ($2.3 \%$), and $90$ kbar ($3.5 \%$).
The QSA values (within V-ZSISA) are instead 
$68$ kbar ($0.9 \%$), $11$ kbar ($0.5 \%$), 
$-5$ kbar ($-0.2 \%$), $50$ kbar
($0.6 \%$), $27$ kbar ($1 \%$) more distant from the experimental data. Moreover, V-ZSISA QSA 
predicts a decrease of $C_{12}$ contrary to experiment.
V-ZSISA QHA predicts an increase of both $C_{12}$ and
$C_{13}$, in qualitative agreement with experiment. 

Figs.~\ref{fig:ec_pressure}, \ref{fig:c12_pressure}, and \ref{fig:c13_pressure} illustrate the pressure dependence of the ECs at select temperatures: $4$ K, $301$ K, and $1000$ K. We examined the pressure range from $0$ kbar to $150$ kbar, which is readily achievable in temperature-dependent high-pressure experiments. Within this range, the ECs exhibit a near-linear relationship with pressure at all temperatures investigated. Notably, $C_{11}$, $C_{33}$, and $C_{44}$ decrease with increasing temperature, whereas $C_{12}$ and $C_{13}$ increase. The $4$ K values are compared with theoretical PBE data from Ref.~\cite{deng_elastic_2009}, demonstrating reasonable agreement. Differences of $112$ kbar for $C_{12}$ and $100$ kbar for $C_{13}$ are visible in Figs.~\ref{fig:c12_pressure} and \ref{fig:c13_pressure}, respectively. While a $148$ kbar difference for $C_{11}$ is present, it is not discernible within the scale of Figure~\ref{fig:ec_pressure}. In contrast, the differences for $C_{33}$ and $C_{44}$ are considerably smaller, at $4$ kbar and $25$ kbar, respectively.

\section{Conclusions}

We have applied to osmium the computational workflow recently introduced for the calculation of the TDECs in hcp metals.~\cite{gong_high-temperature_2024} We presented both the QSA and the QHA results obtained using the PBE functional. 
Experimental data were available for comparison only within the temperature range $5-301$ K. We observed that, consistent with previous findings, the QHA data exhibit closer agreement with experimental results than the QSA calculations.

We evaluated the zero static internal strain approximation (ZSISA) and found it to have negligible impact on the calculated $C_{11}$ and 
$C_{12}$ ECs. Furthermore, in the QSA calculations, we examined the V-ZSISA. This approximation proved to be negligible on a large scale, but manifesting as a discernible difference on the enlarged scale employed for comparison with experimental data. Nevertheless, the QSA calculations appear to provide a reasonable quantification of the V-ZSISA effect.

The QHA results, derived via interpolation on the stress-pressure curve (within V-ZSISA), were compared with direct calculations performed at the geometry corresponding to $1500$ K. Observed discrepancies were of a magnitude comparable to those encountered between the QSA V-ZSISA and full-grid interpolation. Consequently, a comprehensive full-grid interpolation of the QHA ECs for osmium, necessitating phonon calculations at $1050$ distorted geometries ($7\times 5\times 30$), is deemed unnecessary at this moment, despite the potential for minor corrections to the V-ZSISA QHA results.

Finally, we have presented the high-pressure, high-temperature ECs of osmium, with the anticipation that these data may serve as a valuable reference for future experimental measurements of these quantities.

\begin{acknowledgments}
    
This work has been supported by the Italian MUR (Ministry of University and Research) through the National Centre for HPC, Big Data, and Quantum Computing (grant No. CN00000013). Computational facilities have been provided by SISSA through its Linux Cluster, ITCS, and the SISSA-CINECA 2021-2025 Agreement. Partial support has been received from the European Union through the MAX ``MAterials design at the eXascale" Centre of Excellence for Supercomputing applications (Grant agreement No. 101093374, co-funded by the European High Performance Computing joint Undertaking (JU) and participating countries 824143).  

\end{acknowledgments}

% \nocite{*}

% \bibliographystyle{apsrev4-1}
\bibliography{apssamp}% Produces the bibliography via BibTeX.

\end{document}